\documentstyle[preprint,prl,aps]{revtex}
\begin{document}
\draft
\title{Helicity Modulus and Meissner Effect in a Fluctuating
   Type II Superconductor}
\author{Tao Chen and S. Teitel}
\address{Department of Physics and Astronomy, University of Rochester,
 Rochester, New York 14627}
\date{\today}
\maketitle
\begin{abstract}
The helicity modulus for a fluctuating type II superconductor is computed
within the elastic medium approximation, as a probe of
superconducting phase coherence and the
Meissner effect in the mixed state.  We argue that at the vortex
line lattice melting transition, there remains
superconducting coherence parallel to the applied magnetic field,
provided the vortex line liquid retains
a finite shear modulus at finite wavevector.
\end{abstract}
\narrowtext

In the high temperature superconductors, fluctuation effects
are believed to be important over a wide region of the $H-T$ phase
diagram\cite{Nel,Fish,Bul,moore,Glaz}.
Recently
there has been much controversy concerning the effect of
vortex line fluctuations on
long range order of the superconducting wavefunction
in the mixed state\cite{moore,Glaz,Hou2,Moo2,Ikeda}.
Part of this controversy has concerned the definition of the
proper gauge invariant correlation function.
In this work we reconsider the question of superconducting coherence
by considering instead the behavior of
the helicity modulus\cite{JasFis} of the superconductor, which
has proven a valuable criteria for
coherence in related superfluid\cite{Ohta} and
Josephson array\cite{TJ} models.
As will be seen below, it is an intrinsically
gauge invariant quantity.  Furthermore, the helicity modulus is
equivalent to the linear response coefficient between an applied
perturbation in magnetic field and the resulting supercurrent.
Hence it more directly probes one of the most characteristic properties
of a superconductor, the partial Meissner effect of the mixed
state.

Working in the elastic medium approximation,
we compute the helicity modulus to
lowest order in the fluctuation of vortex lines.
We show that vortex line fluctuations have a dramatically
different effect on the helicity modulus parallel versus
transverse to the applied magnetic field.
For the parallel case, the helicity modulus gives
total screening as in the Meissner effect.
We argue that this total screening is unaffected by the
vortex line lattice melting transition, provided the vortex liquid
retains a finite shear modulus at finite wavevector.
In the limit of an extreme type II
superconductor with $\lambda\to\infty$,
our results explain recent numerical simulations by
Li and Teitel\cite{LT} which show clearly the persistence of
phase coherence parallel to the applied field, well into the
vortex liquid state.
Similar results were
obtained by Feigel'man $et$ $al.$\cite{Fei}, working with a related
$2d$ boson model.  We show a simple relation between their
results and ours, which can be expressed in terms of the
``winding number" of vortex lines.  For simplicity, we carry
out our calculation for an isotropic superconductor; the
extension to the uniaxial anisotropic case is straightforward.
We work within the London approximation, which should be valid
provided one is not too close to $H_{c2}$.

The Landau-Ginzburg Helmholtz free energy\cite{Tink}
for an isotropic uniform superconductor, within the London
approximation of constant wavefunction amplitude, can be written as,
\begin{equation}
{\cal H} ={1\over 2}J_0\int d^3r \left\{ |{\bf\nabla}\theta -{\bf A}|^2
          + \lambda^2 |{\bf\nabla}\times {\bf A}|^2\right\}
\label{eq:htheta}
\end{equation}
where $\theta$ is the phase of the superconducting wavefunction,
$\lambda$ is the ``bare" magnetic penetration length,
$J_0=\phi_0^2/16\pi^3\lambda^2$ with $\phi_0$ the flux quantum,
and $(\phi_0/2\pi){\bf A}$ is the magnetic vector potential.
${\bf \nabla}\times {\bf A}= 2\pi {\bf f}$ where
${\bf f}({\bf r})\equiv {\bf B}({\bf r})/\phi_0$ is the local density of
magnetic flux quanta.
The partition function $Z$ is computed averaging over
{\it independently}\cite{note1} fluctuating $\theta$ and ${\bf A}$,
subject to
the constraint that $\langle {\bf f}({\bf r})\rangle =
f{\bf \hat z}$ for a uniform average magnetic induction $B{\bf \hat z}$.
In evaluating
Eq.(\ref{eq:htheta}), the integration is to be cut off at the core
of a vortex line, so that the energy stays finite.

In terms of the supervelocity ${\bf v}\equiv {\bf\nabla}\theta -{\bf A}$,
and its Fourier transform ${\bf v}_q \equiv \int d^3r \,{\rm e}^{i{\bf q}
\cdot {\bf r}} {\bf v}({\bf r})$, the helicity modulus
is defined\cite{Ohta} as the linear response coefficient
between induced supercurrent $J_0{\bf v}$ and an applied twist in phase.
If we take ${\bf v}_q \to {\bf v}_q + \delta v_q{\rm\bf \hat\mu}$ in
Eq.(\ref{eq:htheta}), then the helicity modulus in direction
${\rm\bf \hat \mu}$ is,
\begin{equation}
\Upsilon_\mu({\bf q}) \equiv {\partial^2 {\cal F}\over\partial
(\delta v_q)^2}\bigg|_{\delta v_q=0}
=J_0\left\{ 1-{J_0\over VT}\langle v_{q\mu}
v_{-q\mu}\rangle\right\}\label{eq:heli}
\end{equation}
where ${\cal F}=-T\ln Z$ is the total free energy, and $V=L_zL_\perp^2$
is the system volume.
Because of the symmetry with which ${\bf \nabla}\theta$ and ${\bf A}$
enter ${\bf v}$, $\Upsilon_\mu$ equivalently
gives the induced supercurrent
that flows in response to an applied perturbation in magnetic field,
given by the
vector potential $\delta {\bf A}_q = \delta v_q{\rm\bf \hat\mu}$.
In evaluating
$\Upsilon_\mu({\bf q})$, the physically relevant case is the
limit $q_\mu\to 0$.  This follows from
the convention of Baym\cite{Baym}, where to describe a system
with a current flowing in the direction ${\rm\bf \hat\mu}$,
the appropriate
thermodynamic limit is to take the system size $L_\mu\to\infty$ first,
followed
by $L_{\nu}\to\infty$ for the directions ${\rm\bf \hat\nu}\perp
{\rm\bf \hat\mu}$ (equivalently, ${\bf q}\cdot\delta {\bf A}_q=0$ in the
London gauge).

Defining the vortex line density ${\bf n}$ by
${\bf \nabla}\times{\bf \nabla}\theta=2\pi {\bf n}$, one can write
an arbitrary configuration
${\bf v}_q$ in gauge invariant form as,
\begin{equation}
{\bf v}_q = 2\pi i\left\{{\bf q}\chi_q +
{{\bf q}\times({\bf n}_q-{\bf f}_q)
\over q^2}\right\}.\label{eq:vq}
\end{equation}
$\chi$ is a smooth function which gives the
longitudinal part of ${\bf v}_q$, while the
transverse part of ${\bf v}_q$ is determined by
${\bf \nabla}\times {\bf v}=2\pi ({\bf n}-{\bf f})$.
Substituting Eq.(\ref{eq:vq}) into the Hamiltonian (\ref{eq:htheta}),
and decoupling the ${\bf n}_q$ and ${\bf f}_q$ degrees of freedom by
completing the square in ${\bf f}_q$, we get
\begin{equation}
{\cal H} = {2\pi^2J_0\over VT}\sum_q\left\{q^2\chi_q\chi_{-q} +
{1\over q^2}(1+\lambda^2q^2)\delta{\bf f}_q\cdot\delta{\bf f}_{-q}
+{\lambda^2\over 1+\lambda^2 q^2}{\bf n}_q\cdot {\bf n}_{-q}\right\}
\label{eq:h2}
\end{equation}
where $\delta{\bf f}_q\equiv {\bf f}_q-{\bf f}_q^0$ is the fluctuation
of the magnetic flux density away from the value
${\bf f}_q^0 = {\bf n}_q/(1+\lambda^2q^2)$, which minimizes
the Hamiltonian for a given vortex configuration ${\bf n}_q$.
The partition sum is now an average over all smooth
functions $\chi$, all $\delta{\bf f}_q$ such that
${\bf q}\cdot \delta{\bf f}_q=0$ (so that
${\bf \nabla}\cdot {\bf B}=0$), and all
vortex configurations ${\bf n}$.  The vortex line
interaction in Eq.(\ref{eq:h2}) is just the familiar London
result\cite{Tink,Bra2}.

The helicity modulus is evaluated by substituting Eq.(\ref{eq:vq})
into Eq.(\ref{eq:heli}), and using the Hamiltonian (\ref{eq:h2}) to
evaluate the averages over $\chi_q$ and $\delta {\bf f}_q$.  Taking
the limit $q_\mu\to 0$ one gets,
\begin{equation}
\lim_{q_\mu\to 0}
\Upsilon_\mu({\bf q}) =
\lim_{q_\mu\to 0} {J_0\lambda^2q^2\over 1+\lambda^2q^2}\left\{ 1
-{4\pi^2J_0\lambda^2\over VT}\, {({\rm\bf\hat\mu}
\times {\bf\hat q})_\alpha
({\rm\bf\hat\mu}\times{\bf\hat q})_\beta \langle n_{q\alpha}
n_{-q\beta}\rangle\over 1+\lambda^2q^2}
\right\}.\label{eq:heli2}
\end{equation}

For a superfluid or spin model\cite{JasFis,Ohta,TJ} phase coherence is
indicated by a non-vanishing
$\Upsilon_\mu$ in the limit ${\bf q}\to 0$.
For the superconductor however, the gauge field ${\bf A}$
is free to adjust itself to screen out the applied phase
twist (or perturbation $\delta {\bf A}$), and so  even in
the superconducting state $\Upsilon_\mu
({\bf q}\to 0)\sim q^2$, as seen in Eq.(\ref{eq:heli2})
above\cite{note2}.
In fact, if no vortex lines are present (${\bf n}_q=0$),
Eq.(\ref{eq:heli2}) just gives the familiar total screening
response of the Meissner state\cite{Baym}.
With the presence of vortex lines in the mixed state, we can
generalize the form of the Meissner response, by defining
a renormalized coupling $(J\lambda^2)_R$ and penetration
length $\lambda_R$ such that
\begin{equation}
\lim_{q\to 0}\Upsilon_\mu\equiv {(J\lambda^2)_Rq^2\over 1+\lambda_R^2q^2}
\label{eq:helir}
\end{equation}
where
$\lambda_R$ and $(J\lambda)_R$ may depend on the
direction ${\bf\hat q}$.
Thus to examine superconductivity
it is necessary to consider the form of
$\Upsilon_\mu$ at small but finite ${\bf q}$.

At high $T$, one can make a hydrodynamic approximation\cite{Mar}
and average over ${\bf n}({\bf r})$ as if it was a continuous
function, subject to the constraint that vorticity is conserved
${\bf q}\cdot {\bf n}_q=0$.  Using the Hamiltonian (\ref{eq:h2})
one gets $\Upsilon_\mu({\bf q}) =0$ as expected.  At low $T$, one
can evaluate the vortex line correlations using the elastic approximation
for small fluctuations about a vortex line lattice.  If ${\bf u}_i(z)$
is the transverse displacement of vortex line $i$ at height $z$ from its
position ${\bf R}_i$ in the vortex line lattice, then
${\bf n}({\bf r}_\perp,z)=\sum_i
{\bf\delta}({\bf r}_\perp -{\bf R}_i-{\bf u}_i(z))
({\bf\hat z} +d{\bf u}_i/dz)$.  To evaluate Eq.(\ref{eq:heli2})
to lowest order in $T$, it is only necessary to consider the
expansion of ${\bf n}_q$ to linear order in ${\bf u}_i$.
For small $q>0$ we have,
\begin{equation}
{\bf n}_q=if({\bf q}\cdot
{\bf u}_q{\bf \hat z} - q_z{\bf u}_q)\label{eq:n2}
\end{equation}
where ${\bf r}_\perp$, ${\bf R}_i$, and ${\bf u}_i$ lie in the $xy$
plane, $q_z$ and ${\bf q}_\perp$ are the components of ${\bf q}$
parallel and perpendicular to ${\bf\hat z}$, and
${\bf u}_q\equiv f\sum_i\int dz\, {\rm e}^{i(q_zz+{\bf q}_\perp
\cdot {\bf R}_i)}{\bf u}_i(z)$.
Correlations of ${\bf u}_q$
may be evaluated using the elastic Hamiltonian,
as derived by Brandt\cite{Bra2},
\begin{equation}
{\cal H}_{el}={1\over 2V}\sum_q\left\{(c_{44}q_z^2 +
c_{11}q_\perp^2)u_{qL}
u_{-qL} + (c_{44}q_z^2 + c_{66}q_\perp^2)
u_{qT}u_{-qT}\right\}
\label{eq:hel}
\end{equation}
where $u_{qL}$ and $u_{qT}$ are the components of ${\bf u}_q$
parallel and transverse to ${\bf q}_\perp$,
and $c_{44}({\bf q})$,
$c_{11}({\bf q})$,
and $c_{66}({\bf q})$ are the tilt, compression, and
shear moduli respectively.

Substituting Eq.(\ref{eq:n2}) into Eq.(\ref{eq:heli2}), and evaluating
the displacement correlations
using ${\cal H}_{el}$, we find for perpendicular and parallel responses,
\begin{equation}
\lim_{q_x\to 0}\Upsilon_x({\bf q}) = \lim_{q_x\to 0}\,
{J_0\lambda^2q^2\over
1+\lambda^2q^2}\left\{1-{B^2\over 4\pi (1+\lambda^2q^2)}{q^2\over
(c_{44}q_z^2+c_{11}q_\perp^2)}\right\}\label{eq:helix}
\end{equation}
\begin{equation}
\lim_{q_z\to 0}\Upsilon_z({\bf q}) = \lim_{q_z\to 0}\,
{J_0\lambda^2q^2\over
1+\lambda^2q^2}\left\{1-{B^2\over 4\pi (1+\lambda^2q^2)}{q_z^2\over
(c_{44}q_z^2 + c_{66}q_\perp^2)}\right\}.\label{eq:heliz}
\end{equation}

For the transverse response
$\Upsilon_x({\bf q})$ there are two
cases to consider: ($i$) ${\bf q}=q{\bf\hat z}$, and
($ii$) ${\bf q}=q{\bf\hat y}$.  In ($i$) the perturbation
$\delta {\bf A}_q$ gives a magnetic induction along
${\bf\hat y}$, oscillating in the ${\bf\hat z}$ direction.
It is thus a tilt modulation of the
original induction $B{\bf\hat z}$.  In ($ii$), the perturbation
gives a magnetic induction along ${\bf\hat z}$,
which oscillates along ${\bf\hat y}$; it is thus
a compression modulation of $B{\bf\hat z}$.  Accordingly,
Eq.(\ref{eq:helix}) shows that in
($i$) $\Upsilon_x$ depends on $c_{44}$, while in ($ii$)
$\Upsilon_x$ depends on $c_{11}$.
We consider in detail case ($i$).
A comparison of Eq.(\ref{eq:helix}) with Eq.(\ref{eq:helir}) shows that
$(J\lambda^2)_R$ is determined by  $c_{44}({\bf q}= 0)$, while
$\lambda_R$ is determined by $dc_{44}(0)/dq^2$.
Using the result of Brandt\cite{Bra2},
\begin{equation}
c_{44}= {B^2\over 4\pi}\left[ {1\over 1+\lambda^2 q^2} +
\left({dH_\perp\over dB_\perp}-1\right)\right]
\label{eq:c11}
\end{equation}
we find
\begin{equation}
{(J\lambda^2)_R\over J_0\lambda^2} = 1-{dB_\perp\over dH_\perp},
\qquad{\lambda_R^2\over\lambda^2} = 1-{dB_\perp\over dH_\perp}.
\label{eq:JR}
\end{equation}
For an isotropic system, the factor $dH_\perp/dB_\perp$ in $c_{44}$
above, where the derivative is evaluated at the average
magnetic induction $B{\bf\hat z}$, is
equal to the more familiar $H/B$.
The renormalization factor for the coupling $(J\lambda^2)_R$
has a simple physical interpretation.
Since the induced magnetic induction
is determined from Maxwell's equations as
${\bf A}_{ind}=-(\Upsilon_x({\bf q})/J_0\lambda^2 q^2)\delta {\bf A}$,
Eq.(\ref{eq:JR}) results in
a fraction $dB_\perp/dH_\perp$ of the perturbation
$\delta H{\bf\hat y}={\bf\nabla}\times\delta {\bf A}$
penetrating the superconductor, while
the remainder is screened out as in the Meissner effect.

We now consider the parallel response $\Upsilon_z$ of
Eq.(\ref{eq:heliz}). As long as the shear modulus
$c_{66}$ is finite in the limit $q_z\to 0$, the
term in Eq.(\ref{eq:heliz})
due to vortex line fluctuations vanishes, and
one has total screening of the perturbation as in the Meissner state.
If $c_{66}$ is identically
zero,  $\Upsilon_z({\bf q}_\perp)$ has exactly the same form
as $\Upsilon_x(q{\bf\hat z})$, with the greatly reduced
coupling $1-dB_\perp/H_\perp$.  A related
analysis of order parameter correlations\cite{Ikeda} has
led Ikeda {\it et al.}
to conclude that vortex line lattice melting (where $c_{66}(0)\to 0$)
should be accompanied by a dramatic
reduction in phase correlations
along the direction of the magnetic field.  However a careful
analysis of Eq.(\ref{eq:heliz}) shows that $\Upsilon_z$
continues to be
unaffected by vortex line fluctuations provided that
$c_{66}(q_z,{\bf q}_\perp)$ does not vanish as fast (or faster than)
$q_z^2$, as $q_z\to 0$ for $finite$ ${\bf q}_\perp$.

One may question the application of elastic theory once
the vortex line lattice has melted.  A possible justification has
been given by Marchetti and Nelson\cite{MarNel3}, who show that a
hexatic
vortex line liquid may be described by an elastic theory
in which one includes free dislocation loops.
Averaging over dislocations,
they find that the elastic moduli $c_{11}$ and $c_{44}$ are largely
unchanged, however the shear
modulus is renormalized to $c_{66}(q_z=0,{\bf q}_\perp)\sim q_\perp^2$.
Thus, while $c_{66}({\bf q}=0)= 0$ as expected for a liquid,
$c_{66}$ remains finite for finite ${\bf q}_\perp$.
Hence Eq.(\ref{eq:heliz}) continues to give total
screening of the perturbation
upon melting of the vortex line lattice, ie. superconducting coherence
persists parallel to the applied field for some finite temperature
range into the vortex liquid state.

Continuing the expansion as in Eq.(\ref{eq:n2}) to next order in
the displacements ${\bf u}_q$, we find a correction only to
$\lambda_R$ of order $\lambda_R^2/\lambda^2\sim
(3.8T/\pi J_0)\sqrt{B/\phi_0}$ (using $B\simeq 0.2H_{c2}$).
Evaluating at the vortex line lattice melting temperature,
which we find to be
$T_M\sim 1.7c_L^2\pi J_0\sqrt{\phi_0/B}$ (where $c_L\sim 0.15$ is the
Lindemann parameter), we find a small correction to $\lambda_R^2$ of
order 15\%.  Thus the conclusions above from the lowest order expansion
continue to hold\cite{note5}.

If one considers the above calculation in the limit of an extreme type II
superconductor where $\lambda\to\infty$, and all fluctuations of the
gauge field ${\bf A}$ are frozen out, we return to the case
analogous to an ordinary superfluid.
Taking the $\lambda\to\infty$ limit in
Eqs.(\ref{eq:helix}$-$\ref{eq:c11}), we find that
$\Upsilon_z({\bf q}\to 0)=J_0$ is finite, and hence the system has
phase
coherence in the ${\bf \hat z}$ direction, even in the vortex line
liquid state (provided $c_{66}> 0$ for ${\bf q}_\perp\ne 0)$.
$\Upsilon_x$ however vanishes at all temperatures, as $B=H$ when
$\lambda\to\infty$.  This explains the recent
numerical results of Li and Teitel\cite{LT}
in a lattice $\lambda\to\infty$ model.
There $\Upsilon_z$ was found to vanish at a $T_{cz}$,
well into the vortex line liquid state, while $\Upsilon_x$
vanished at a much lower $T_{c\perp}$, where the vortex line lattice
melted.  The finite $\Upsilon_x$ for this model at low $T$
is due to the effects of pinning introduced by the
discrete numerical mesh, which creates a finite energy
barrier to small $q$ elastic distortions.  This effectively
adds a $q$ independent constant to the denominator of the second
term in Eq.(\ref{eq:helix}), so that as $q\to 0$, $\Upsilon_x=J_0$.
Only when the vortex line lattice melts will thermal fluctuations
dominate over pinning, and one recovers Eq.(\ref{eq:JR})
with the resulting $\Upsilon_x=0$.

Much work has been done using
an analogy between fluctuating vortex lines, and the imaginary time
world lines of two dimensional bosons\cite{Nel,Fei}.
Feigel'man et $al$.\cite{Fei}
have used this analogy to argue that in the large
$\lambda$ limit, there will exist a boson normal fluid phase intermediate
between the boson lattice and the boson superfluid phases.  They
predict that the corresponding phase of the superconductor is
characterized by
coherence parallel to the applied magnetic field,
but not transverse to it.
It is interesting to examine this analogy within the above
elastic approximation.  A convenient
expression for the $2d$ boson superfluid density has been given
by Ceperley and Pollock\cite{CP} in terms of the
``winding number" ${\bf W}$ of boson world lines,
$\rho_s=mT\langle W^2\rangle/2\hbar^2$.
If there are $only$ magnetic field induced vortex
lines fluctuating in a directed fashion\cite{note4}
(ie. a single valued displacement ${\bf u}_i(z)$),
\begin{equation}
\langle W^2\rangle = {1\over L_\perp^2}\langle\big|
\sum_i[{\bf u}_i(L_z)-{\bf u}_i(0)]\big|^2\rangle=
{1\over L_\perp^2}\langle {\bf n}_{q=0}^\perp\cdot
{\bf n}_{-q=0}^\perp\rangle\label{eq:rhos}
\end{equation}
where ${\bf n}_{q=0}^\perp$ is the average vortex line density
transverse to the average magnetic induction $B{\bf\hat z}$.
Translating\cite{Nel} from $2d$ bosons to vortex lines
($\hbar/T_{boson}\to L_z$, $\hbar\to T_{vortex}$,
$m\to\epsilon_1\sim\pi J_0$
the single vortex line tension), and
evaluating $\langle W^2\rangle$ within the elastic approximation,
one finds\cite{note3}
\begin{equation}
\rho_s = \lim_{q_\perp\to 0}\,
\lim_{q_z\to 0} \,{\epsilon_1 f^2\over 2}\left\{
{q_z^2\over c_{44}q_z^2 +c_{66}q_\perp^2}
+{q_z^2\over c_{44}q_z^2 +c_{11}q_\perp^2}\right\}.
\label{eq:rhos2}
\end{equation}
The second term above always vanishes when one takes $q_z\to 0$ first,
as $c_{11}$ is always finite.
The first term is just the same factor as appears in Eq.(\ref{eq:heliz})
for $\Upsilon_z$.  Hence in the vortex line liquid, if
$c_{66}({\bf q}_\perp\ne 0)> 0$,
we have both total Meisnner screening of perturbations
$\delta A_{q_\perp}{\bf\hat z}$, and $\rho_s=0$, consistent
with the predictions of Feigel'man $et$ $al$.

When $c_{66}$
vanishes identically, Eq.(\ref{eq:rhos2}) gives
$\rho_s=\epsilon_1 f^2/2c_{44}(0)$.
This result also follows from a direct evaluation of Eq.(\ref{eq:rhos})
within a hydrodynamic approximation\cite{Mar}.
If the $2d$ boson normal to superfluid transition is of
the Kosterlitz-Thouless type, then the universal jump in $\rho_s$
at the transition may be written\cite{CP} as
$\langle W^2_c\rangle=4/\pi$.  This gives a transition temperature
for the vortex lines, $T_{KT}=(\phi_0^2/\pi^2 L_z)dH_\perp/dB_\perp$.
For $H\simeq H_{c1}$,
this result gives $T_{KT}\sim 1/B$ in good agreement with
an earlier estimate
by Nelson\cite{note6}.  At larger $B$, where
$dH_\perp/dB_\perp\simeq 1$,
$T_{KT}$ is independent of $B$.
For sufficiently large $L_z$ however, this
KT transition is presumably preempted by the transition to the
hexatic vortex line liquid, in which $c_{66}({\bf q}_\perp)>0$.

We would like to thank Profs. E. Domany,
D. R. Nelson, A. Schwimmer, and especially M. Feigel'man and P. Muzikar,
for very helpful discussions.
S. T.
wishes to thank the hospitality of the Weizmann Institute of Science
where this work was begun, and BSF grant 89-00382 which made that visit
possible.  This work has been supported by U. S. Department
of Energy grant DE-FG02-89ER14017.


\begin{references}
\bibitem{note2} If we did not average over
 $\delta{\bf f}_q$, ie. if we followed the procedure of
 Refs.\cite{Glaz,Hou2} (see \cite{note1} above), we would
 find that $\Upsilon_\mu({\bf q\to 0})>0$
 for $all$ $T$,  clearly a non-physical result.
\bibitem{note5} For an anisotropic system with
  $\gamma\equiv\lambda_z/\lambda_\perp$,
  this higher order correction to $\lambda_R^2/\lambda_z^2$
  increases by a factor $\gamma$, while $T_M$ decreases by a
  factor $\gamma$.  Thus the correction at the melting transition
  remains the same as in the isotropic case.
\bibitem{note3} The limit $q_z\to 0$ in Eq.(\ref{eq:rhos2})
  is not meant to imply
  that we are only interested in the $L_z\to\infty$ case, but
  only to indicate the correct way to obtain the ${\bf q}=0$ correlation
  which defines $W^2$ in Eq.(\ref{eq:rhos}).  That this is the
  correct order for the limits is seen by comparing the result
  of Eq.(\ref{eq:rhos2}) when $c_{66}\equiv 0$, with the direct
  evaluation of Eq.(\ref{eq:rhos}) in the high $T$ hydrodynamic
  approximation\cite{Mar}.
\bibitem{note6} To see that this result agrees with that of
  Ref.\cite{Nel},
  note that near $H_{c1}$, $dH_\perp/dB_\perp\simeq H_{c1}/\gamma^2 B$
  where $\gamma\equiv\lambda_z/\lambda_\perp$ gives the anisotropy.
\end{references}
\end{document}